\documentclass[aps,prb,reprint,superscriptaddress,showpacs]{revtex4-1}
\usepackage{amsmath,amssymb}
\usepackage{amsbsy}
\usepackage{graphicx}

\newcommand{\abs}[1]{\left| #1 \right|}
\newcommand{\vecn}{\mathbf{n}}
\newcommand{\jtr}{\mathbf{j}_{\mathrm{tr}}}

\newcommand{\vecd}{\mathbf{d}}
\newcommand{\vecrho}{\boldsymbol{\rho}}
\newcommand{\vecl}{\mathbf{L}}
\newcommand{\vecg}{\mathbf{g}}
\newcommand{\psionei}{\psi_1^{(i)}}
\newcommand{\him}{h_{\mathrm{im}}}

\begin{document}

\title{Pinning of an Abrikosov vortex on a small cylindrical cavity: A Ginzburg-Landau approach}
\author{A. A. Bespalov}
\affiliation{Institute for Physics of Microstructures,
 Russian Academy of Sciences, GSP-105, 603950, Nizhny Novgorod, Russia}
\affiliation{Univ. Bordeaux, LOMA, UMR 5798, F-33600 Talence, France}
\author{A. S. Mel'nikov}
\affiliation{Institute for Physics of Microstructures,
 Russian Academy of Sciences, GSP-105, 603950, Nizhny Novgorod, Russia}
\begin{abstract}
Within the Ginzburg-Landau theory we consider Abrikosov vortex pinning on a columnar defect with the characteristic size of the cross-section $D$ much smaller than the coherence length $\xi$. We present an extension of the electrostatic analogies method, which proved to be useful for calculations of the pinning force for large cavities ($D \gg \xi$), to the case of small defects ($D \ll \xi$). The pinning potential for an elliptic cavity is derived analytically. Also, we determine the depinning current for a circular defect.
\end{abstract}

\pacs{74.25.Wx, 74.20.De}

\maketitle

\section{Introduction}

The possibilities for practical applications of type-II superconductors depend crucially on the quality of vortex pinning structures, which can be embedded into these materials. Columnar defects proved to be the most efficient pinning centers. Nowadays, various techniques exist allowing to create disordered arrays of such defects,\cite{Civale+91} as well as regular defect lattices.\cite{Welp++}

The first theoretical study of vortex pinning on a cylindrical defect has been carried out by Mkrtchyan and Schmidt.\cite{Mkrtchyan+71} In their paper the London equation has been solved exactly for a vortex interacting with a cavity in the form of a circular cylinder. The pinning force has been analyzed in detail for the cavity radius $a$ satisfiyng the condition $\xi \ll a \ll \lambda$, where $\xi$ is the coherence length and $\lambda$ is the London length. Later,\cite{Nordborg+2000} this analysis has been extended to the case of large cavities with $a \gg \lambda$. Buzdin and Feinberg\cite{Buzdin+96} pointed out that London screening can be neglected in a large range of fields in extreme type-II superconductors. This observation allowed them to establish an electrostatic analogy and to simplify considerably the solution for a vortex interacting with a circular cavity: it has been demonstrated that the full magnetic field can be presented as the sum of the own vortex field and the field of image vortices, situated inside the cavity. Also, using the conformal transformation technique, pinning potentials for more tricky columnar cavities have been derived.\cite{Buzdin+98,Buzdin+2000} However, in the calculation of the pinning potential for non-circular defects only the field of the image vortices has been transformed, while the modification of the own field of the real vortex has not been taken into account.

In order to analyze small defects, a more complex approach is required. For temperatures close to the superconducting critical temperature the Ginzburg-Landau (GL) approximation is a natural choice. It has been applied for the analysis of vortex pinning on columnar defects of different shape and nature. Blatter at al.\cite{Blatter+94} estimated the pinning potential for a small defect using a simple variational procedure, taking into account only the suppression of the order parameter inside the defect. Maurer et al.\cite{Maurer+96} calculated numerically the pinning energy for a vortex centered on a circular insulating or metallic inclusion. In Ref. \onlinecite{Priour+2003} vortex interaction with a cylindrical hole with a square cross-section has been analyzed numerically. In  Ref. \onlinecite{Rosenstein+2010} the critical current for a vortex lattice pinned on a set of defects with reduced critical temperature has been determined using a variational procedure and numerical simulations. Yet, exact analytical solutions of the GL equation have been lacking so far.

In this paper, within the GL theory we consider the interaction of a vortex with a small cylindrical cavity or insulating inclusion with the characteristic size of the cross-section $\xi_0 \ll D \ll \xi$, where $\xi_0$ is the zero-temperature coherence length (in the case $D \ll \xi_0$ the correct description can be obtained only on the basis of a microscopic theory\cite{Thuneberg+82,Melnikov+2009}). We present the exact pinning potentials in terms of the unperturbed vortex order parameter for a circular and elliptic defect. For the circular cavity the depinning current is also determined. For the treatment of a small defect with arbirary cross-section we propose an electrostatic analogy, which serves as a counterpart to the mentioned above analogy between the London theory (valid for large defects, $D \gg \xi$) and electrostatics. Finally, using the conformal transformation method developed in Ref. \onlinecite{Buzdin+2000}, we derive the pinning potential for an elliptic cavity within the London approximation, taking into account the modification of both the own vortex field and the image field.

\label{sec:Introduction}

\section{Vortex pinning within the Ginzburg-Landau theory}
\label{sec:GL}

\subsection{Basic equations}
\label{sub:basic}

The starting point for our analysis is the GL equation for the order parameter $\psi = \abs{\psi}e^{i\theta}$:
\begin{equation}
	-\xi^2 \nabla^2 \psi - \psi + n_0^{-1} \abs{\psi}^2 \psi = 0,
	\label{eq:GL}
\end{equation}
where $n_0$ is the concentration of Cooper pairs in the bulk. The GL parameter is assumed to be large, so the vector potential can be neglected.\cite{Gor'kov+75} Let us put the origin of coordinates inside the insulating defect, or cavity, and the $z$-axis along the vortex axis. Then if the vortex axis is parallel to the generatrix of the defect and perpendicular to the transport current, the order parameter does not depend on $z$. This is the case that will be analyzed further.

Equation \eqref{eq:GL} is supplemented by two boundary conditions, specifying the normal derivative of $\psi$ at the insulating defect border and the transport current density $\jtr$ far from the vortex core:
\begin{equation}
	\vecn \nabla \psi  \biggl|_{\partial S} = 0,
	\label{eq:bound_a}
\end{equation}
\begin{equation}
	\frac{2e \hbar \abs{\psi}^2 \nabla \theta}{m} \biggl|_{\rho \rightarrow \infty}= \jtr.
	\label{eq:bound_far}
\end{equation}
Here $S$ is the defect cross-section, $\partial S$ denotes the border of $S$, $\vecn$ is the outward unit normal to $\partial S$, $e$ is the electron charge, and $m$ is the Cooper pair mass.

Since our system must contain one vortex, an additional condition for the order parameter phase arises:
\begin{equation}
	 \oint \nabla \theta d\mathbf{l} = 2 \pi,
	 \label{eq:theta_condition}
\end{equation}
where integration is performed over a sufficiently large contour surrounding the defect.

We expect that for some current $j_d$ Eqs. \eqref{eq:GL} - \eqref{eq:theta_condition} can be solved when $j_{\mathrm{tr}}<j_d$, and no solution exists when $j_{\mathrm{tr}}>j_d$. Then it is natural to consider $j_d$ as a depinning current.

\begin{figure}[t]
  \includegraphics[scale=1]{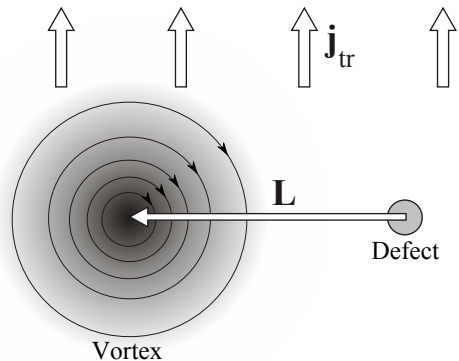}
  \caption{The cross-section of the system: in the presence of a transport current $\jtr$ the vortex is shifted by a vector $\vecl$ with respect to the defect.}\label{fig:1}
\end{figure}

We will solve Eqs. \eqref{eq:GL} - \eqref{eq:theta_condition} in the case $j_{\mathrm{tr}}<j_d$. If the transport current is much smaller than the depairing current, the order parameter has the following asymptotics at infinity:
\begin{equation}
	\psi = \sqrt{n_0} e^{i\varphi + i\mathbf{q} \vecrho} + O(\rho^{-1}),
	\label{eq:asympt}
\end{equation}
where $\varphi$ is the polar angle, and $\mathbf{q} = m \jtr/2e\hbar n_0$. This asymptotics can be derived from Eqs. \eqref{eq:GL}, \eqref{eq:bound_far} and \eqref{eq:theta_condition} if one expands $\psi$ in powers of $\rho^{-1}$ and neglects terms proportional to $j_{\mathrm{tr}}^2$. Now we make some assumptions concerning the behavior of the order parameter in the vicinity of the defect.
\begin{itemize}
	\item[\bf (A)] $\psi(\boldsymbol{\rho})$ reaches its asymptotic behavior at sufficiently small distances from the origin: $\psi \approx \sqrt{n_0} e^{i \varphi} \cdot e^{i \mathbf{q} \boldsymbol{\rho}}$ when $\rho \geq R$, where $R$ is some radius in the range $\xi \ll R \ll q^{-1}$.
	\item[\bf (B)] The vortex is weakly distorted by a small defect and a small current. This means that the solution of Eqs. \eqref{eq:GL} - \eqref{eq:theta_condition} can be presented in the form $\psi=\psi_0 + \psi_1$, where $\psi_0$ corresponds to an unperturbed vortex shifted from the origin by a vector $\vecl$ (see Fig. \ref{fig:1}), and $\psi_1$ is a small perturbation: $\abs{\psi_1(\rho)} \ll \sqrt{n_0}$ when $\rho \ll q^{-1}$.
\end{itemize}
The assumption (B) is justified by the fact that the unperturbed vortex corresponds to a local minimum of the free energy, so large distortions are not energetically favorable. Both statements (A) and (B) can be verified by numerical calculations.

Let us write down the equations for the function $\psi_1$. If we linearize Eq. \eqref{eq:GL} we obtain
\begin{equation}
	-\xi^2 \nabla^2 \psi_1 - \psi_1 + 2 n_0^{-1} \abs{\psi_0}^2 \psi_1 + n_0^{-1}\psi_0^2 \psi_1^* = 0.
	\label{eq:psi_1}
\end{equation}
According to the statement (A),
\begin{equation}
	\psi_1(\boldsymbol{\rho}) = \psi(\boldsymbol{\rho}) - \psi_0(\boldsymbol{\rho}) \approx \sqrt{n_0}i e^{i \varphi} (\mathbf{q} \boldsymbol{\rho}), \quad R<\rho \ll q^{-1}.
	\label{eq:bound_psi1_far}
\end{equation}
The boundary condition at the defect border for $\psi_1$ follows from Eq. \eqref{eq:bound_a}:
\begin{equation}
	(\nabla \psi_1 + \nabla \psi_0) \mathbf{n} \biggl|_{\partial S} = 0.
	\label{eq:bound_psi_1}
\end{equation}
Thus, Eqs. \eqref{eq:psi_1} - \eqref{eq:bound_psi_1} are to be solved.

\subsection{Variational derivation of the pinning potential}
\label{sub:simple}

In this section, for the reader's convenience, we present a relatively simple, but not rigorous derivation of the pinning potential. A more detailed and careful analysis is given in section \ref{sub:balance}.

We will determine the free energy $F$ per unit length of a vortex shifted from the origin by a vector $\vecl$:
\[ F = \frac{H_c^2}{4\pi n_0} \int_{\vecrho \notin S} \left( \xi^2 \abs{\nabla \psi}^2 - \abs{\psi}^2 + n_0^{-1} \frac{\abs{\psi}^4}{2} \right) d^2 \vecrho, \]
where $H_c$ is the thermodynamic critical field satisfying the relation
\[ \frac{H_c^2}{8\pi} = \frac{\hbar^2 n_0}{4m \xi^2}. \]
In the zero-order approximation $F$ equals the free energy of an unperturbed vortex:
\[ F_0 = \frac{H_c^2}{4\pi n_0} \int \left( \xi^2 \abs{\nabla \psi_0}^2 - \abs{\psi_0}^2 + n_0^{-1} \frac{\abs{\psi_0}^4}{2} \right) d^2 \vecrho. \]
The pinning potential equals the difference between the exact free energy $F$ and $F_0$: $U_p(\vecl) = F(\vecl) - F_0$. This difference consists of two terms -- $\Delta F_1$ and $\Delta F_2$. The first term is connected with the suppression of the order parameter inside the defect:
\begin{equation}
	\Delta F_1 \approx -S \frac{H_c^2}{4\pi n_0} \left( \xi^2 \abs{\nabla \psi_0}^2 - \abs{\psi_0}^2 + n_0^{-1} \frac{\abs{\psi_0}^4}{2} \right) \Biggl|_{\vecrho=0}.
	\label{eq:DF_1}
\end{equation}
Note that this expression is valid only for very small defects with the characteristic size of the cross-section $D \ll \xi$. The second term is connected with the distortion of the order parameter outside the defect:
\begin{eqnarray*}
 & \Delta F_2 = \frac{H_c^2}{4\pi n_0} \int_{\vecrho \notin S} \left( \xi^2 \abs{\nabla \psi}^2 - \abs{\psi}^2 + n_0^{-1} \frac{\abs{\psi}^4}{2} \right) d^2 \vecrho & \\
 & - \frac{H_c^2}{4\pi n_0} \int_{\vecrho \notin S} \left( \xi^2 \abs{\nabla \psi_0}^2 - \abs{\psi_0}^2 + n_0^{-1} \frac{\abs{\psi_0}^4}{2} \right) d^2 \vecrho. &
\end{eqnarray*}
We substitute here $\psi = \psi_0 + \psi_1$, where $\psi_1$ is a small perturbation satisfying Eqs. \eqref{eq:psi_1} and \eqref{eq:bound_psi_1} in the vicinity of the defect and decaying to zero on a scale $\rho \sim \xi$ (if we use the function $\psi_1$ satisfying Eq. \eqref{eq:psi_1} in the whole $xy$ plane, the component $\Delta F_2$ will diverge to positive infinity). Close to the cavity the characteristic length scale for $\psi_1$ is $D \ll \xi$, hence, we can neglect terms of the order of $\abs{\psi_1}^2$ as compared to the term $\xi^2 \abs{\nabla \psi_1}^2$:
\begin{eqnarray*}
	& \Delta F_2 = \frac{\xi^2 H_c^2}{4\pi n_0} \left\{ \int_{\vecrho \notin S} \abs{\nabla \psi_1}^2 d^2 \vecrho \right. & \\
	& \left. - \int_{\partial S} \left[ \psi_1^*(\nabla \psi_0 \vecn) + \psi_1 (\nabla \psi_0^* \vecn) \right] d\ell \right\} &
\end{eqnarray*}
Using Eq. \eqref{eq:bound_psi_1} and applying the Gauss theorem, we transform the right-hand side of the last relation as follows:
\begin{equation}
	 \Delta F_2 = -\frac{\xi^2 H_c^2}{4\pi n_0} \int_{\vecrho \notin S} \left( \abs{\nabla \psi_1}^2 + \psi_1^* \nabla^2 \psi_1  + \psi_1 \nabla^2 \psi_1^* \right) d^2 \vecrho
	\label{eq:intermediate}
\end{equation}
From Eq. \eqref{eq:psi_1} we find that 
\[ \abs{\psi_1 \nabla^2 \psi_1} \sim \abs{\psi_1}^2/\xi^2 \ll \abs{\nabla \psi_1}^2, \]
hence
\begin{equation}
	\Delta F_2 \approx -\frac{\xi^2 H_c^2}{4\pi n_0} \int_{\vecrho \notin S} \abs{\nabla \psi_1}^2 d^2 \vecrho.
	\label{eq:F_2}
\end{equation}
An explicit expression for $\Delta F_2$ in terms of $\psi_0$ will be given below (see Eq. \eqref{eq:F_2'}).

\subsection{Force balance equation}
\label{sub:balance}

In this section we derive the solvability condition for the system \eqref{eq:psi_1} - \eqref{eq:bound_psi_1}. Our derivation closely follows the computations from Ref. \onlinecite{Gor'kov+75} which were used to determine the viscous drag force acting on a moving vortex.

First, we introduce the auxiliary function $\psi_d = \vecd \nabla \psi_0$, where $\vecd$ is an arbitrary constant unit vector. $\psi_d$ satifies the equation
\begin{equation}
	-\xi^2 \nabla^2 \psi_d - \psi_d + 2 n_0^{-1} \abs{\psi_0}^2 \psi_d + n_0^{-1} \psi_0^2 \psi_d^* = 0.
	\label{eq:psi_d}
\end{equation}
Let us multiply \eqref{eq:psi_1} by $\psi_d^*$ and subtract Eq. \eqref{eq:psi_d} multiplied by $\psi_1^*$ from it. When we add the complex conjugate to the resulting equation we obtain
\[ \mathrm{div} \left( -\psi_d^* \nabla \psi_1 + \psi_1 \nabla \psi_d^* - \psi_d \nabla \psi_1^*  + \psi_1^* \nabla \psi_d \right) = 0. \]
We integrate the last relation over the region $\vecrho \notin S$, $\rho < R$ and apply the Gauss theorem:
\begin{eqnarray}
	& \int_{\rho = R} \left( -\psi_d^* \nabla \psi_1 + \psi_1 \nabla \psi_d^* - \psi_d \nabla \psi_1^* + \psi_1^* \nabla \psi_d \right) \vecn_1 d\ell & \nonumber \\ 
	& - \int_{\partial S} \left( -\psi_d^* \nabla \psi_1 + \psi_1 \nabla \psi_d^* -  \psi_d \nabla \psi_1^* + \psi_1^* \nabla \psi_d \right) \vecn d\ell = 0, & \nonumber \\
	&&
	\label{eq:integrals}
\end{eqnarray}
where $\vecn_1$ is the outward unit normal to the circle $\rho=R$. The first integral can be calculated with the help of Eq.\eqref{eq:bound_psi1_far}:
\begin{eqnarray}
		& \int_{\rho = R} \left( -\psi_d^* \nabla \psi_1 + \psi_1 \nabla \psi_d^* - \psi_d \nabla \psi_1^*  + \psi_1^* \nabla \psi_d \right) \vecn d\ell  & \nonumber \\
		& \approx - \frac{2 \pi m}{e \hbar}\left (\vecd ; \left[ \mathbf{z}_0 ; \jtr \right] \right), &
		\label{eq:far_int}
\end{eqnarray}
where $\mathbf{z}_0$ is the unit vector along the $z$ axis.

The second integral in Eq. \eqref{eq:integrals} can be transformed using Eq. \eqref{eq:bound_psi_1} and the Gauss theorem:
\begin{eqnarray}
& \int_{\partial S} \left( -\psi_d^* \nabla \psi_1 -  \psi_d \nabla \psi_1^* + \psi_1 \nabla \psi_d^*  + \psi_1^* \nabla \psi_d \right) \vecn d\ell & \nonumber \\
& \approx S \cdot \mathrm{div} \left( \psi_d^* \nabla \psi_0 + \psi_d \nabla \psi_0^* \right) \biggl|_{\vecrho = 0} & \nonumber \\
&  + \int_{\partial S} \left( \psi_1 \nabla \psi_d^*(0)  + \psi_1^* \nabla \psi_d(0) \right) \vecn d\ell. & 
\label{eq:def_int}
\end{eqnarray}
Here and further we neglect terms which are much smaller than $n_0 D^2/\xi^3$.

In order to proceed we have to determine the value of $\psi_1$ at the defect boundary. It has been noted in Sec. \ref{sub:simple} that $\psi_1 \sim \sqrt{n_0}D\xi^{-1}$. Hence, 
\[ \partial^2 \psi_1 / \partial x^2 \sim \partial^2 \psi_1 / \partial y^2 \sim \sqrt{n_0}D^{-1} \xi^{-1} \gg \abs{\psi_1} \xi^{-2}. \]
This implies that near the defect we can use the Laplace equation
\begin{equation}
	\nabla^2 \psi_1 = 0.
	\label{eq:Laplace}
\end{equation}
instead of Eq. \eqref{eq:psi_1}. The boundary condition can also be simplified:
\begin{equation}
	\nabla \psi_1 \mathbf{n} \biggl|_{\partial S} = - \nabla \psi_0 (0) \mathbf{n}.
	\label{eq:bound_psi_1'}
\end{equation}
Such simplification is acceptable since we are not interested in small corrections of the order of $\sqrt{n_0}D^2/\xi^2$ to $\psi_1$. Equations \eqref{eq:Laplace} and \eqref{eq:bound_psi_1'} are equivalent to an electrostatic problem where $\psi_1$ plays the role of the electric potential of a charged cylinder. Note that these equations can not be derived within the electrostatic approximation for the London theory,\cite{Buzdin+96} where variations of the superconducting phase are taken into account, but the order parameter modulus is assumed to be constant. 

The relation
 \[ \oint_{\partial S} \nabla \psi_1 \vecn d\ell = 0, \]
can be interpreted as a vanishing total ``charge'' of the cylinder. It provides that a solution of Eqs. \eqref{eq:Laplace} and \eqref{eq:bound_psi_1'} exists that decays like $\rho^{-1}$ at infinity. This solution, which we denote as $\psi_1^{(d)}$, represents the irregular part of $\psi_1$: it has singularities inside the defect. We define the regular component of $\psi_1$ as
\begin{equation}
	\psi_1^{(i)} = \psi_1 - \psi_1^{(d)}.
	\label{eq:psi1i_def}
\end{equation}
It is proved in the Appendix that the contribution of $\psi_1^{(i)}$ to the integral in the right-hand side of Eq. \eqref{eq:def_int} is negligible. 

Combining Eqs. \eqref{eq:integrals} - \eqref{eq:def_int} and using the fact that $\psi_0$ satisfies Eq. \eqref{eq:GL} we obtain
\begin{eqnarray}
& \frac{2\pi \xi^2 m}{e\hbar} \left (\vecd  \left[ \mathbf{z}_0 ; \jtr \right] \right)  +  S (\vecd  \nabla) \! \left( \xi^2  \abs{\nabla \psi_0}^2 \!\! - \abs{\psi_0}^2 \!\! + \frac{ \abs{\psi_0}^4}{2n_0} \right) \! \Biggl|_{\vecrho = 0}  & \nonumber \\
& + \xi^2 \int_{\partial S} \left( \psi_1^{(d)} \nabla \psi_d^*(0)  + \psi_1^{(d)*} \nabla \psi_d(0) \right) \vecn d\ell \approx 0.&
\label{eq:balance1}
\end{eqnarray}
Owing to the linearity of Eqs. \eqref{eq:Laplace} and \eqref{eq:bound_psi_1'} the solution can be presented in the form
\begin{equation}
	\psi_1^{(d)} = \left( \vecg(\vecrho); \nabla \psi_0(0) \right),
	\label{psi1_solution}
\end{equation}
where $\vecg$ is  a real vector field defined by the relations
\begin{equation}
	\nabla^2 \vecg =0, \qquad	(\vecn \nabla) \vecg  \biggl|_{\partial S} = - \vecn, \qquad	\vecg \biggl|_{\rho \to \infty} = 0.
	\label{eq:vecg}
\end{equation}
Then
\begin{eqnarray}
 & \int_{\partial S} \left(\psi_1^{(d)} \nabla \psi_d^*(0) + \psi_1^{(d)*} \nabla \psi_d(0) \right) \vecn d\ell & \nonumber \\
 & = (\vecd \nabla) \left( \nabla \psi_0 \hat{G} \nabla \psi_0^* \right) \biggl|_{\vecrho=0}, & \nonumber
\end{eqnarray}
where $\hat{G}$ is a real symmetric matrix with components
\begin{equation}
	G_{ij} = \int_{\partial S} g_i n_j d\ell = \int_{\vecrho \notin S} \nabla g_i \nabla g_j \, d^2 \vecrho.
	\label{G_ij}
\end{equation}
Equation \eqref{eq:balance1} transforms into
\[ -\frac{\phi_0}{c} \left (\vecd  \left[ \mathbf{z}_0 ; \jtr \right] \right) - (\vecd \nabla_{\vecl}) U_p= 0. \]
Here $\phi_0 = \pi \hbar c/e$ is the flux quantum, $\nabla_{\vecl} = \partial/\partial \vecl$, and
\begin{eqnarray}
	& U_p = - S \frac{H_c^2}{4\pi n_0} \left( \xi^2 \abs{\nabla \psi_0}^2  - \abs{\psi_0}^2 \! + \frac{\abs{\psi_0}^4}{2n_0} \right)  \Biggl|_{\vecrho = 0} & \nonumber \\
	& -  \xi^2 \frac{H_c^2}{4\pi n_0} \nabla \psi_0(0) \hat{G} \nabla \psi_0^*(0). &
	\label{eq:U_p}
\end{eqnarray}
Since $\vecd$ is an arbitrary vector, it can be dropped, and we finally obtain the force balance equation, connecting the vortex displacement $\vecl$ with the transport current $\jtr$:
\begin{equation}
	-\frac{\phi_0}{c} \left[ \mathbf{z}_0 ; \jtr \right] - \nabla_{\vecl} U_p= 0.
	\label{eq:balance}
\end{equation}
Here, the first term is the Lorentz force and the second term is the pinning force: $F_p =- \nabla_{\vecl} U_p$. Thus, we may conclude that $U_p$ is the pinning potential. It can be proved that this definition of the pinning potential is identical to the one given in Sec. \ref{sub:simple}, if $\psi_1$ in Eq. \eqref{eq:F_2} is replaced by $\psi_1^{(d)}$. Thus,
\begin{equation}
	\Delta F_2 = -  \xi^2 \frac{H_c^2}{4\pi n_0} \nabla \psi_0(0) \hat{G} \nabla \psi_0^*(0).
	\label{eq:F_2'}
\end{equation}
Before we determine some pinning potentials explicitly, we would like to note that our consideration can be easily generalized for the anisotropic case. Indeed, generally, the GL free energy can be presented in the form
\begin{eqnarray}
	& F = \frac{H_c^2}{4\pi n_0} \int \left( \xi_x^2 \abs{\frac{\partial \psi}{\partial x}}^2 + \xi_y^2 \abs{\frac{\partial \psi}{\partial y}}^2 + \xi_z^2 \abs{\frac{\partial \psi}{\partial z}}^2  \right. & \nonumber \\
	& \left. - \abs{\psi}^2 + \frac{\abs{\psi}^4}{2n_0} \right) d^3 \mathbf{r},&
	\label{eq:anisotropic}
\end{eqnarray}
where $\xi_x$, $\xi_y$ and $\xi_z$ are the coherence lengths for different directions. The scaling transformation $\tilde{x} = x$, $\tilde{y} = y \xi_x/ \xi_y$, and $\tilde{z} = z \xi_x / \xi_z$ reduces the free energy to the isotropic form. Thus, we again arrive at Eqs. \eqref{eq:GL} - \eqref{eq:bound_far}.

Now we consider two types of defects.

\subsection{A circular defect}
\label{sub:circular}

Let the defect be a circular cylinder with the radius $a$. When the origin is placed on the axis of the cylinder, the decaying solution of Eqs. \eqref{eq:Laplace} and \eqref{eq:bound_psi_1'} is
\begin{equation}
	\psi_1^{(d)} = \frac{a^2 (\nabla \psi_0; \vecrho)}{\rho^2},
	\label{eq:psi1d_circ}
\end{equation}
and the pinning potential is
\begin{equation}
	 U_p(\vecl)  = -\frac{H_c^2 a^2}{4 n_0} \left( 2 \xi^2 \abs{\nabla \psi_0}^2 - \abs{\psi_0}^2 + \frac{\abs{\psi_0}^4}{2n_0} \right) \Biggl|_{\vecrho = 0}.
	 \label{eq:Up_circ}
\end{equation}
The function $\psi_0$ can be determined numerically from the GL equation \eqref{eq:psi_1}. We are not going into details of these calculations here. A detailed numerical analysis of this function can be found in Ref. \onlinecite{Hu72}.

Equation \eqref{eq:Up_circ} allows us to determine the pinning energy, $E_p$, and compare it with a numerical result from  preceding paper. According to our calculations,
\[ E_p = U_p(\infty) - U_p(0) = 0.47 H_c^2 a^2, \]
which coincides with the numerical value given in Ref. \onlinecite{Maurer+96} up to a factor of the order of unity.

\begin{figure}[t]
  \includegraphics[scale=1.35]{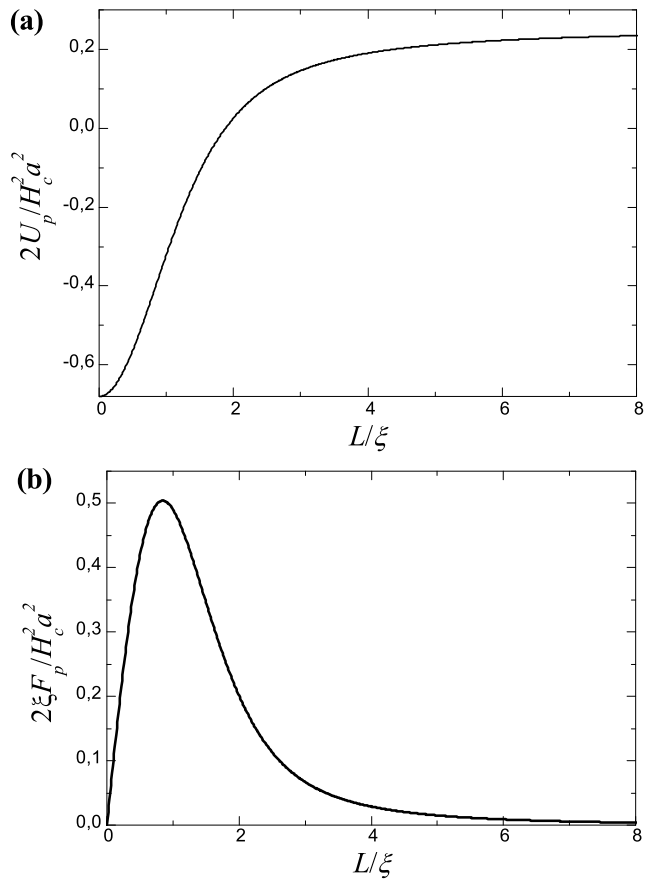}
  \caption{Profiles of the pinning potential (a) and the pinning force (b) for a circular defect.}\label{fig:2}
\end{figure}

The profiles of the pinning potential and the pinning force are plotted in Fig. \ref{fig:2}. The pinning force reaches its maximum at $L=L_{\mathrm{cr}}=0.84\xi$, where $F_p = F_{\mathrm{cr}} = 0.252 H_c^2 a^2/\xi$. When $j_{\mathrm{tr}}> c F_{\mathrm{cr}}/\abs{\phi_0}$, the force balance equation \eqref{eq:balance} has no solutions, hence
\begin{equation}
	j_d = c F_{\mathrm{cr}}/\abs{\phi_0} = 0.252 \frac{H_c^2 a^2 \abs{e}}{\pi \hbar \xi}
	\label{eq:j_d_circ}
\end{equation}
is the depinning current. When $j<j_d$ Eq. \eqref{eq:balance} has two solutions due to the nonmonotonic behavior of the function $F_p(L)$, but the solution with the larger vortex displacement is thermodynamically unstable. Indeed, it can be easily proved that it corresponds to a saddle point of the correction $\Delta F$ to the free energy of a vortex connected with the presence of the defect and the transport current:
\[ \Delta F = U_p + \frac{\phi_0}{c} \left[ (\mathbf{z}_0 \times \jtr) \vecrho \right]. \]

To sum up, deppining from a circular defect occurs as follows: as the transport current increases from zero to $j_d$ the vortex displacement with respect to the origin increases from zero to $L_{\mathrm{cr}}$; as the current is increased further the vortex is carried away from the pinning cite.

\subsection{An elliptic defect}
\label{sub:elliptic}

Consider an elliptic defect with the cross-section
\[ \frac{x^2}{a^2} + \frac{y^2}{b^2} < 1, \]
where $a>b$. We shall determine the vector field $\vecg$ and the pinning potential. It is convenient to use the elliptic coordinates $(\zeta,\eta)$:
\[ x = \sqrt{a^2 - b^2} \cosh{\zeta} \cos{\eta}, \qquad y = \sqrt{a^2 - b^2} \sinh{\zeta} \sin{\eta}. \]
The border of the defect corresponds to the value $\zeta = \zeta_0$, where
\[ \sinh{\zeta_0} = \frac{b}{\sqrt{a^2 - b^2}}, \qquad \cosh{\zeta_0} = \frac{a}{\sqrt{a^2 - b^2}}. \]
Eqs. \eqref{eq:vecg} in the new coordinates read
\begin{eqnarray}
	& \frac{\partial^2 \vecg}{\partial \zeta^2} + \frac{\partial^2 \vecg}{\partial \eta^2} = 0, \qquad \vecg \biggl|_{\zeta \to \infty} = 0, & \label{eq:g_elliptic1} \\
	& \frac{\partial g_x}{\partial \zeta} \biggl|_{\zeta = \zeta_0} = -b \cos{\eta}, \qquad \frac{\partial g_y}{\partial \zeta} \biggl|_{\zeta = \zeta_0} = -a \sin{\eta}. &
	\label{eq:g_elliptic2}
\end{eqnarray}
The solution is
\begin{equation}
	g_x = b e^{\zeta_0 - \zeta} \cos{\eta}, \qquad g_y = a e^{\zeta_0 - \zeta} \sin{\eta}.
	\label{eq:g_elliptic_sol}
\end{equation}
Using Eq. \eqref{G_ij}, we obtain the components of the matrix $\hat{G}$:
\begin{equation}
	G_{xx} = \pi b^2, \quad G_{yy} = \pi a^2, \quad G_{xy} = G_{yx} = 0.
	\label{eq:G_elliptic}
\end{equation}
Hence, according to Eq. \eqref{eq:U_p}, the pinning potential is
\begin{eqnarray}
	& U_p = - \frac{H_c^2}{4n_0} \left[ ab \left( \xi^2 \abs{\nabla \psi_0}^2 - \abs{\psi_0}^2 + \frac{\abs{\psi_0}^4}{2n_0} \right) \right. & \nonumber \\
	& \left. +  \xi^2 b^2 \abs{\frac{\partial \psi_0}{\partial x}}^2 + \xi^2 a^2 \abs{\frac{\partial \psi_0}{\partial y}}^2 \right] \Biggl|_{\vecrho = 0}. & 
	\label{eq:U_p_elliptic}
\end{eqnarray}
The potential well for the vortex now does not have cylindrical symmetry. As a result, the vortex displacement $L$ and the depinning threshold $j_d$ depend on the direction of the transport current.

\section{Comparison with the London theory.}
\label{sec:London}

In a preceding paper\cite{Buzdin+98} the pinning potential in the presence of a circular and elliptic cavity has been derived within the London theory. The applicability condition for this approach is that the defect size be much larger than the temperature-dependent coherence length, i.e., $D \gg \xi(T)$, so it may seem that the results from Ref. \onlinecite{Buzdin+98} can not be compared with ours. However, the condition $D \ll \xi$ is not essential for our calculations. Indeed, instead we may demand (i) $\abs{\psi_1} \ll \sqrt{n_0}$ and (ii): the quantity $\nabla \psi_0$ should be approximately constant in the area occupied by the cavity. These two conditions are satisfied when
\begin{equation}
	D \gg \xi \mbox{    and    } L \gg D,
	\label{eq:far+large}
\end{equation}
so for a large defect and large vortex-defect distance our pinning potential should coincide with the one obtained within the London theory.

For a circular defect our pinning potential \eqref{eq:Up_circ} at large vortex-defect distances, $L \gg a$, equals
\begin{equation}
	U_p = -\left( \frac{\phi_0}{4\pi \lambda} \right)^2 \frac{a^2}{L^2} + \mathrm{const}.
	\label{eq:Circ_dimensional}
\end{equation}
This expression is in good agreement with Eq. (5) from Ref. \onlinecite{Buzdin+98}. For an elliptic hole Eq. \eqref{eq:U_p_elliptic} yields in the $L \gg a$ limit
\begin{equation}
	U_p = -\left( \frac{\phi_0}{4\pi \lambda} \right)^2 \frac{1}{2} \left(\frac{ab}{L^2} + \frac{b^2 L_y^2}{L^4} + \frac{a^2 L_x^2}{L^4} \right) + \mathrm{const},
	\label{eq:Elliptic_dimensional}
\end{equation}
whereas the potential from Ref. \onlinecite{Buzdin+98} is
\begin{equation}
	U_p = - \left( \frac{\phi_0}{4\pi \lambda} \right)^2 \left( \frac{a+b}{2} \right)^2 \frac{1}{L^2},
	\label{eq:Ellipse_Buzdin}
\end{equation}
which, obviously, does not coincide with \eqref{eq:Elliptic_dimensional}. Below we will explain the reason of this discrepancy.

The derivation of the interaction energy between a vortex and a cavity in the London approximation is based on the equation
\begin{equation}
	U_p = \frac{\phi_0 \him (\vecl)}{8\pi},
	\label{eq:Up_London}
\end{equation}
where $\him$ is the $z$ projection of the field created by image vortices. This field can be expressed as
\begin{equation}
	\him = h_z - \frac{\phi_0}{2\pi \lambda^2} \ln{\abs{\frac{\lambda^2}{\zeta - \zeta_0}}},
	\label{eq:h_im}
\end{equation}
where $\zeta = x+iy$, $\zeta_0 = L_x + iL_y$, and $h_z$ is the full magnetic field, satisfying the Poisson equation
\begin{equation}
	\nabla^2 h_z = -\frac{\phi_0}{\lambda^2} \delta(\vecrho-\vecl).
	\label{eq:Poison}
\end{equation}
The second term in the right-hand side of Eq. \eqref{eq:h_im} represents the own field of the vortex with the opposite sign. For a circular defect with the radius $a_0$ the image field at the position  of the vortex is
\begin{equation}
	\him^c(\zeta_0) = \frac{\phi_0}{2\pi \lambda^2} \ln\left( 1 - \frac{a_0^2}{\abs{\zeta_0}^2} \right),
	\label{eq:h_im_circ}
\end{equation}
In order to obtain the magnetic field in the presence of a non-circular defect, we may apply a conformal transformation $w=w(\zeta)$ to the $\zeta$ plane. Since the form of Poisson's equation is not modified by such a transformation, the field distribution in the $w$ plane is given by
\[	h_z(w) =h_z^c(\zeta(w)), \]
where $h_z^c(\zeta)$ is the solution of Eq. \eqref{eq:Poison} in the presence of a circular defect. Using the definition \eqref{eq:h_im} of the image field, we obtain
\begin{eqnarray}
	& \him(w_0) = \him^c(\zeta(w_0)) & \nonumber \\
	& + \left[  \frac{\phi_0}{2\pi \lambda^2} \ln{\abs{\frac{\lambda}{\zeta(w) - \zeta_0}}} - \frac{\phi_0}{2\pi \lambda^2} \ln{\abs{\frac{\lambda}{w - w_0}}} \right] \biggr|_{w=w_0}, &
	\label{eq:him_correct}
\end{eqnarray}
where $w_0 = w(\zeta_0)$ specifies the position of the vortex in the $w$-plane. Hence, the pinning potential equals
\begin{equation}
	U_p = \left(\frac{\phi_0}{4\pi \lambda} \right)^2 \left[ \ln\left( 1 - \frac{R^2}{\abs{\zeta(w_0)}^2} \right) - \ln \abs{ \frac{d \zeta}{d w} (w_0)} \right].
	\label{eq:U_correct}
\end{equation}
Here, the first logarithmic term originates from the transformation of the image field \eqref{eq:h_im_circ}, while the second term is connected with the modification of the own field of the vortex. In Ref. \onlinecite{Buzdin+98} this term has not been taken into account. As a result, the isotropic potential \eqref{eq:Ellipse_Buzdin} has been obtained. 
In order to determine the correct pinning potential for an elliptic cavity, we apply the modified Joukovski transformation: \cite{Buzdin+98}
\begin{equation}
	w(\zeta) = \frac{a+b}{2} \frac{\zeta}{a_0} + \frac{a-b}{2} \frac{a_0}{\zeta};
	\label{eq:Joukovski}
\end{equation}
\begin{eqnarray}
	& U_p = \left(\frac{\phi_0}{4\pi \lambda} \right)^2 \left[ \ln\left( 1- \abs{\frac{a+b}{w+\sqrt{w^2 - a^2 + b^2}}}^2 \right) \right. & \nonumber \\
	& \left. -\ln\abs{1 + \frac{w}{\sqrt{w^2 - a^2 + b^2}}} \right] + \mathrm{const}. &
	\label{eq:Ellipse_correct}
\end{eqnarray}
For $\abs{w} \gg a$ this expression coincides with our result, obtained within the GL theory (see Eq. \eqref{eq:Elliptic_dimensional}).

\section{Conclusion}

By solving the Ginzburg-Landau equation, we developed a method to determine analytically the pinning potential for a vortex interacting with a small cylindrical cavity. This method has been applied to a circular and elliptic defect. In the latter case, the pinning potential appeared to be anisotropic, as one would expect. Also, we recalculated the pinning potential for an elliptic cavity within the London theory, using the conformal transformation technique,\cite{Buzdin+98} considering the modification of the image field as well as the transformation of the own field of the vortex, which had not been previously taken into account. Our results obtained within the GL and London theories agree well with each other in the range of parameters, where both approaches are valid.

All our previous consideration has been related to the case of a vortex strictly parallel to the defect. This assumption is obviously satisfied in thin superconducting films, where the transport current is distributed almost uniformly over the film thickness. However, it has been claimed\cite{Indenbom+2000} that in a bulk superconductor depinning is likely to occur due to vortex kink formation in a surface layer with the thickness equal to the London length. Still, if the vortex radius of curvature is large as compared to $\xi$, and the vortex axis makes a small angle with the defect axis, our approach should give reasonable estimates of the pinning energy and of the local vortex dispalcement with respect to the defect. Moreover, in the presence of sufficiently small transport current a bound state should occur, when the vortex core is outside the defect (at least in a surface layer -- in bulk superconductors). This bound state should be observable, for example, using scanning tunneling microscopy. 

Our results may be useful for estimations of the depinning current and for manipulations of the critical current anisotropy in superconducting materials.

\section{Acknowledgements}
We are thankful to A. I. Buzdin for helpful discussion.
This work was supported, in part, by European IRSES
program SIMTECH (contract n.246937), the Russian Foundation for
Basic Research, FTP Scientific and educational personnel of
innovative Russia in 2009-2013, and the program of LEA "Physique
Theorique et Matiere Condensee".

\appendix
\section{}
\label{app:psi1i}

In this appendix we will demonstrate that the contribution from the function $\psi_1^{(i)}$ (see Eq. \eqref{eq:psi1i_def}) to the pinning force is negligible. It is sufficient to prove that the absolute value of the integral
\[ I = \int_{\partial S} \left( \psi_1^{(i)} \nabla \psi_d^*(0)  + \psi_1^{(i)*} \nabla \psi_d(0) \right) \vecn d\ell \]
is much smaller than $n_0 D^2/\xi^3$. 

The function $\psi_1^{(i)}$ has the following properties:
\begin{equation}
	\nabla \psi_1^{(i)} \mathbf{n} \biggl|_{\partial S} \approx 0, \qquad \nabla^2 \psi_1^{(i)} = 0, \quad \rho < r,
	\label{eq:psi1i}
\end{equation}
where $r$ is a quantity of the order of the coherence length. Let us introduce an auxiliary function $v$ defined by the relations
\[ \nabla^2 v = 0, \qquad \vecn \nabla v \biggl|_{\partial S} = \xi \vecn \nabla \psi_d^*(0) , \qquad v \biggl|_{\rho \to \infty} = 0. \]
The properties of this function are identical to those of $\psi_1^{(d)}$: it is of the order of $\sqrt{n_0} D \xi^{-1}$ at the defect border and decays like $\rho^{-1}$ at infinity. For a smooth defect $v \sim \sqrt{n_0} D^2/\xi \rho$. Now we make some simple calculations:
\begin{eqnarray}
 & 0 = \int_{\rho \notin S, \, \rho < r} \left (\psionei \nabla^2 v - v \nabla^2 \psionei \right) d^2 \vecrho & \nonumber \\
 & =-\int_{\partial S} \psionei \nabla v \, \vecn d \ell + \int_{\rho=r} \left( \psionei \frac{\partial v}{\partial \rho} - v \frac{\partial \psionei}{\partial \rho} \right) d \ell. & \nonumber
\end{eqnarray}
Since $\psionei \approx \psi_1$ when $\rho \sim r$,
\begin{equation}
	 I = \xi^{-1} \int_{\rho=r} \left( \psi_1 \frac{\partial v}{\partial \rho} - v \frac{\partial \psi_1}{\partial \rho} \right) d \ell + c. c.,
	 \label{eq:I}
\end{equation}
where $c.c.$ denotes the complex conjugate. According to statement (B) from section \ref{sub:basic}, when $\rho=r$ , $\abs{\psi_1} \ll \sqrt{n_0}$ and  $\abs{\partial \psi_1/\partial \rho} \ll \sqrt{n_0}/\xi$, since the characteristic length scale is $\xi$. Then it follows immediately from Eq. \eqref{eq:I} that $\abs{I} \ll n_0 D^2/\xi^3$.

\end{document}